\documentclass[12pt]{article}
\usepackage{scicite}

\usepackage{times}

\usepackage{graphicx}
\usepackage{placeins}
\usepackage{amsmath}
\usepackage{bm}% bold math
\usepackage{color}
\usepackage[normalem]{ulem} % can be removed for submission

\newenvironment{sciabstract}{%
\begin{quote} \bf}
{\end{quote}}

\title{} 
\title{Broadband acoustic invisibility and illusions}
\author
{Theodor S. Becker,$^{1\ast}$ Dirk-Jan van Manen,$^{1}$ \\ 
Thomas Haag,$^{1}$ Christoph B\"arlocher,$^{1}$ Xun Li,$^{1}$ Nele B\"orsing,$^{1}$ \\ 
Andrew Curtis,$^{1,2}$ Marc Serra-Garcia,$^{1}$ Johan O.A. Robertsson$^{1}$\\
\\
\normalsize{$^{1}$Institute of Geophysics, ETH Z\"urich,}\\
\normalsize{8092 Z\"urich, Switzerland}\\
\normalsize{$^{2}$Grant Institute of Geoscience, University of Edinburgh,}\\
\normalsize{Edinburgh, EH9 3FE, United Kingdom}\\
\\
\normalsize{$^\ast$corresponding author; E-mail:  theodor.becker@erdw.ethz.ch.}
}

\date{}

\begin{document} 

% Double-space the manuscript.

\baselineskip24pt

% Make the title.

\maketitle

% Place your abstract within the special {sciabstract} environment.
\begin{sciabstract}
Rendering objects invisible to impinging acoustic waves (cloaking) and creating acoustic illusions (holography) has been attempted using active and passive approaches. While passive methods are applicable only to narrow frequency bands, active approaches attempt to respond dynamically, interfering with broadband incident or scattered wavefields by emitting secondary waves. Without prior knowledge of the primary wavefield, the signals for the secondary sources need to be estimated and adapted in real-time. This has thus far impeded active cloaking and holography for broadband wavefields. We present experimental results of active acoustic cloaking and holography without prior knowledge of the wavefield so that objects remain invisible and illusions intact even for broadband moving sources. This opens novel research directions and facilitates practical applications including architectural acoustics, and stealth.
\end{sciabstract}
\textbf{One Sentence Summary:\\}Creating acoustic invisibility and illusions by replacing parts of a physical medium with a virtual medium through real-time, broadband wavefield control\\

Rendering objects invisible to acoustic, elastic, or electromagnetic waves (cloaking) or making objects appear where there are none (holography) is of immediate scientific and technological interest. Sophisticated cloaks and holograms relying on real-time wavefield manipulation (Fig.~1%\ref{fig:CloakingSchematic}
) remained science-fiction until theoretical advances, particularly in the field of transformation optics \cite{Pendry2006,Leonhardt2006},
%theoretical and technological advances [e.g., transformation optics \cite{Pendry2006,Leonhardt2006} and additive manufacturing \cite{Wong2012}], which
enabled scientists to create the first physical invisibility and illusion devices \cite{Friot2004,Friot2006a,Schurig2006,Ergin2010,Chen2011,Popa2011,Chen2013a}. %Ni2015,Chu2018,Ang2020} 
%and to create wavefield illusions \cite{Kan2013,Melde2016,Zhang2018,Marzo2019}. %Such wavefield phenomena are of immediate interest to many disciplines, including architectural, urban, entertainment and military applications. 
% 
Existing devices rely on the control of wavefields with passive or active methods, but have serious limitations.
% passive methods
Passive methods \cite{Schurig2006,Pendry2006,Ergin2010,Chen2011,Jiang2011,Sanchis2013,Chen2013a,Kan2013,Ni2015,Kadic2016,Kan2016}  %,Wang2019,Li2019c,Zhang2020a}
are based on tailored materials that usually do not adapt to changing incident wavefields, work efficiently only over narrow frequency bands, and often suffer from internal dissipation \cite{Fleury2015}. %Monticone2014b 
% active methods
Active methods \cite{Friot2004,Friot2006a,Miller2006,Vasquez2009,Ma2013,Zhu2014,Selvanayagam2014,Ang2020} rely on the emission of a secondary wavefield that (destructively) interferes with the primary field. The secondary wavefield is typically created with control sources distributed on the boundary or across the area of interest (Fig.~1%\ref{fig:CloakingSchematic}
). Active methods provide a more flexible and dynamic control of wavefields and can, in principle, react to changes in the incident field in real-time. 
% problem with such active methods and some active cloaking literature
However, existing active cloaking and holography devices have considerable limitations: they rely on \textit{a priori} knowledge of the incident or scattered wavefields and hence fail if the characteristics of the primary energy sources change or are unknown \cite{Selvanayagam2014,House2020,Ang2020}%House2020a
, which is the case for many conceivable applications. In such a scenario, the signals for the control sources need to be estimated and updated in real-time. To our knowledge, Ref.~\cite{Friot2006a} are the only authors who have experimentally demonstrated active cloaking of acoustic waves without \textit{a priori} knowledge of the incident wavefield beyond 1D. However, they employed a control algorithm that relies on a non-deterministic optimization approach that reduces the scattered field in certain areas, while it actually \textit{increases} the field in other locations. Moreover, in these experiments the forward-scattered field is not controlled at all, the results are not spectrally broadband, and some knowledge of the signal emitted by the primary source is required. These assumptions compromise the cloak and limit its practical use. %to ensure stability of the underlying control algorithm. 
We consider a fundamentally different approach to acoustic cloaking and holography that relies on virtually replacing part of a physical medium by a desired, virtual medium. This is achieved by complementing the acoustic waves propagating in the physical medium with a real-time simulation of the waves propagating in the virtual medium. The simulation is performed on a massively parallelized, low-latency control system based on field-programmable gate arrays [FPGAs; see \cite{Becker2018} for details]. The two media are linked in real-time through an active, fully deterministic, and global control loop  \cite{ScienceSM,VanManen2007,VanManen2015} that extrapolates particle velocity and pressure wavefields from an acoustically transparent array of microphones to an array of control loudspeakers surrounding the region to be replaced [Fig.~2%\ref{fig:Setup}
; \cite{ScienceSM}]. Using this approach, we demonstrate in two-dimensional (2D) acoustic experiments that a physical scattering object can be replaced with a virtual homogeneous background medium in real-time, thereby hiding the object from broadband acoustic waves (cloaking). In a second set of experiments, part of a physical homogeneous medium is replaced by a virtual scattering object, thereby creating an acoustic illusion of an object that is not physically present (holography). Due to the broadband nature of the control loop and in contrast to other cloaking approaches, no \textit{a priori} knowledge of the primary energy source, nor of the scattered wavefields is required and the approach holds even for primary sources whose locations change over time.

% Results
We first demonstrate cloaking of a quasi-rigid, circular scatterer with a diameter of 12.6~cm placed in the center of an air-filled, 2D acoustic waveguide. The boundary of the scatterer is lined with 20 secondary loudspeakers (Fig.~2D%\ref{fig:Setup}D
). A primary wavefield is emitted by 8 individual loudspeakers located along the perimeter of the experiment % in a radial distance of ~66 cm from the scatterer and 
across an arc of 96$^{\circ}$ (Fig.~3%\ref{fig:ResultsCloakingRigid}
A). All loudspeakers emit the same broadband wavelet [described by the second derivative of a Gaussian function \cite{Wang2015}] with a center frequency, $f_c$, of 3~kHz, but with an increasing time-delay. This mimics a primary source that changes its location with an effective velocity of 173~m/s, sending a broadband pulse from each new location.
First, the control loudspeakers surrounding the scatterer are inactive. Consequently, the rigid scatterer disturbs the incident wavefield causing back- and forward scattering (Fig.~3%\ref{fig:ResultsCloakingRigid}
, A,~D,~G, and J). In a second experiment, the same primary wavefield is emitted, but this time the control loudspeakers complement the scattered field based on the particle velocity field extrapolated in real-time from two circular microphone arrays enclosing the scatterer (Fig.~3%\ref{fig:ResultsCloakingRigid}
,~B,~E,~H, and K). Since the Green's functions used for the extrapolation simulate a homogeneous, air-filled medium, the scattering object is effectively replaced by air through the emission of the secondary wavefield and therefore rendered acoustically invisible for an observer outside of the control loudspeaker array. For comparison, we emit the same primary field, but remove the scattering object and secondary loudspeakers from the waveguide (Fig.~3%\ref{fig:ResultsCloakingRigid}
,~C,~F,~and~I). The resulting acoustic wavefield closely resembles the wavefield obtained with enabled secondary loudspeakers, confirming that the scatterer is indeed rendered acoustically invisible. The scattered field is suppressed not only for the direct wave propagating between the primary source and the scattering object, but also for all echoes from the boundary of the waveguide at later times. The cloak is effective over a broad range of frequencies and all angles with a reduction in mean scattered acoustic intensity of 8.4~dB (Fig.~3%\ref{fig:ResultsCloakingRigid}
,~L~and~M). Note that the scatterer remains acoustically invisible despite the change of location of the primary source, demonstrating that our approach does not rely on any \textit{a priori} knowledge of the incident or scattered fields. Finite-element simulations replicating the three experiments are in excellent agreement with the experimental results (Fig.~3%\ref{fig:ResultsCloakingRigid}
,~G~to~I).\\ 

In a second suite of experiments, we demonstrate the ability to create broadband acoustic holograms. For these experiments, the Green's functions used for the real-time wavefield extrapolation simulate wavefield interactions with virtual objects located within the control loudspeaker array. Since in this case no physical boundary exists at the location of the loudspeakers, wavefields are controlled at a sound-transparent emitting surface. This requires the dual emission (and extrapolation) of particle velocity and pressure as the signatures of collocated monopole and dipole sources \cite{ScienceSM,VanManen2015}.  Consequently, in these experiments, the wavefield is controlled with two closely-spaced circular loudspeaker arrays mounted flush with the planar boundary of the waveguide to create effective dipole and monopole sources [see Fig.~2%\ref{fig:Setup}
E and \cite{ScienceSM}]. We choose extrapolation Green's functions that represent a virtual rigid scatterer with the same geometry as the physical scatterer used in the cloaking experiments (i.e., circular, 12.6~cm diameter, rigid boundary; geometry shown in Fig.~4%\ref{fig:ResultsHolograhpyTransparent}
A). This time, a primary wavefield is created by a single stationary loudspeaker placed at [0,~0.66]~m emitting a wavelet with $f_c~=~3$~kHz. The pressure field is measured by the microphone arrays and extrapolated to the sound-transparent emitting surface in real-time, where the resulting virtually scattered field is emitted by the secondary loudspeakers. As a consequence, a hologram of the virtual scatterer is created: the total wavefield now contains interactions of the primary loudspeaker with the virtual object. The resulting acoustic wavefield and angular distribution of \textit{virtually} scattered intensity are in excellent agreement with measurements made with a \textit{physical} scatterer inside the waveguide (Fig.~4%\ref{fig:ResultsHolograhpyTransparent}
,~B-F), further underlining the successful creation of a broadband acoustic hologram.

We demonstrated the ability to replace part of a physical medium with a virtual medium through real-time and broadband wavefield control. In these experiments, we either replaced a physical scatterer by virtual air, thereby rendering the scatterer acoustically invisible, or created the illusion of a virtual rigid object located within the array of control loudspeakers. The presented methodology opens up novel research in wave physics, particularly because any desired virtual medium of arbitrary complexity, including non-linear or non-physical (e.g., energy-gaining) media, can be inserted into the physical medium within the loudspeaker arrays, as long as the medium can be described by an appropriate set of extrapolation Green's functions. In fact, by physically recording the scattering Green's functions of an existing object and using those Green's functions for the real-time extrapolation in the presented holography approach, physical objects can be acoustically cloned and reproduced in a different environment \cite{ScienceSM}. In principle, the methodology also extends to elastodynamic and electromagnetic waves, but a practical realization is significantly more challenging. The approach has immediate practical implications, including hiding arbitrary objects from interrogating devices, improving architectural acoustics, creating invisible sensors, and creating holograms for communication, educational, entertainment, and stealth purposes. %To give just two examples: 1) the ability to create arbitrarily complex acoustic illusions without the need to physically create such objects presents tremendous opportunities to study detection and imaging methods; 2) individual walls of two physically separated meeting rooms can be acoustically removed, and the two rooms acoustically linked, to create the illusion of being in the same room and sitting at opposite ends of a conference table. 

\section*{Acknowledgments}
%The authors thank four anonymous reviewers for their comments and constructive feedback and Badreddine Assouar for editorial handling. Furthermore, T.S.B. wishes to thank Emily Rose Ciscato, Patrick Elison, Henrik Rasmus Thomsen and Xun Li for discussions, reviewing and feedback.
\textbf{Funding:} This work was funded by the European Research Council (ERC) under the European Union’s Horizon 2020 research and innovation programme (grant agreement No 694407).\\
\textbf{Authors contributions:} Data curation and formal analysis: T.S.B.; Funding acquisition: J.O.A.R. and D.M.; Investigation: T.S.B.; Methodology: D.M., A.C., J.O.A.R., T.S.B., N.B., and X.L.; Resources and software: T.H., C.B., N.B., and T.S.B.; Validation and visualization: T.S.B.; Writing: T.S.B., D.M., J.O.A.R., X.L., A.C., and M.S.\\
\textbf{Competing interests}: The authors declare no competing interests.\\
\textbf{Data and materials availability:} Matlab codes for the analysis, raw experimental data, and numerical simulations will be deposited in an appropriate digital repository. %are available at \textbf{insert link to some archive})

\clearpage

% FIG 1
\begin{figure}[tb!]
	\centering
	\includegraphics{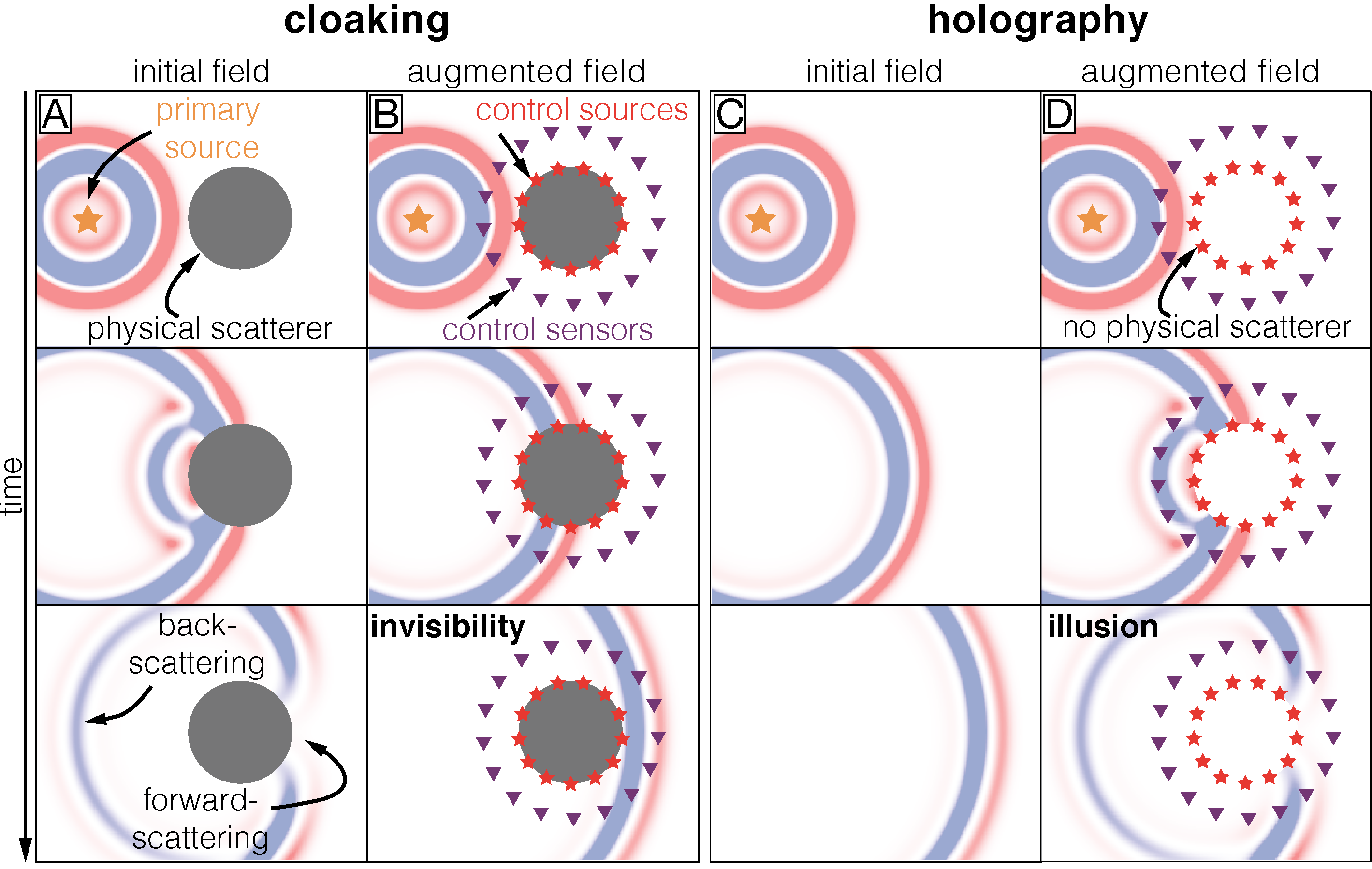}
	\caption{\label{fig:CloakingSchematic}\baselineskip24pt Simulations demonstrating active cloaking and holography. An initial wavefield emitted by a primary source is scattered by an object in the case of cloaking (A) or propagates unobstructed in the case of holography (C). Through active control sources, the incident wavefield is augmented to either cloak the scattering object (B) or create a hologram of an object that is not physically present (D). The input to the control sources is obtained by real-time forward extrapolation using measurements of the control sensors.}
\end{figure}

% FIG 2
\begin{figure}[tb!]
	\centering
	\includegraphics{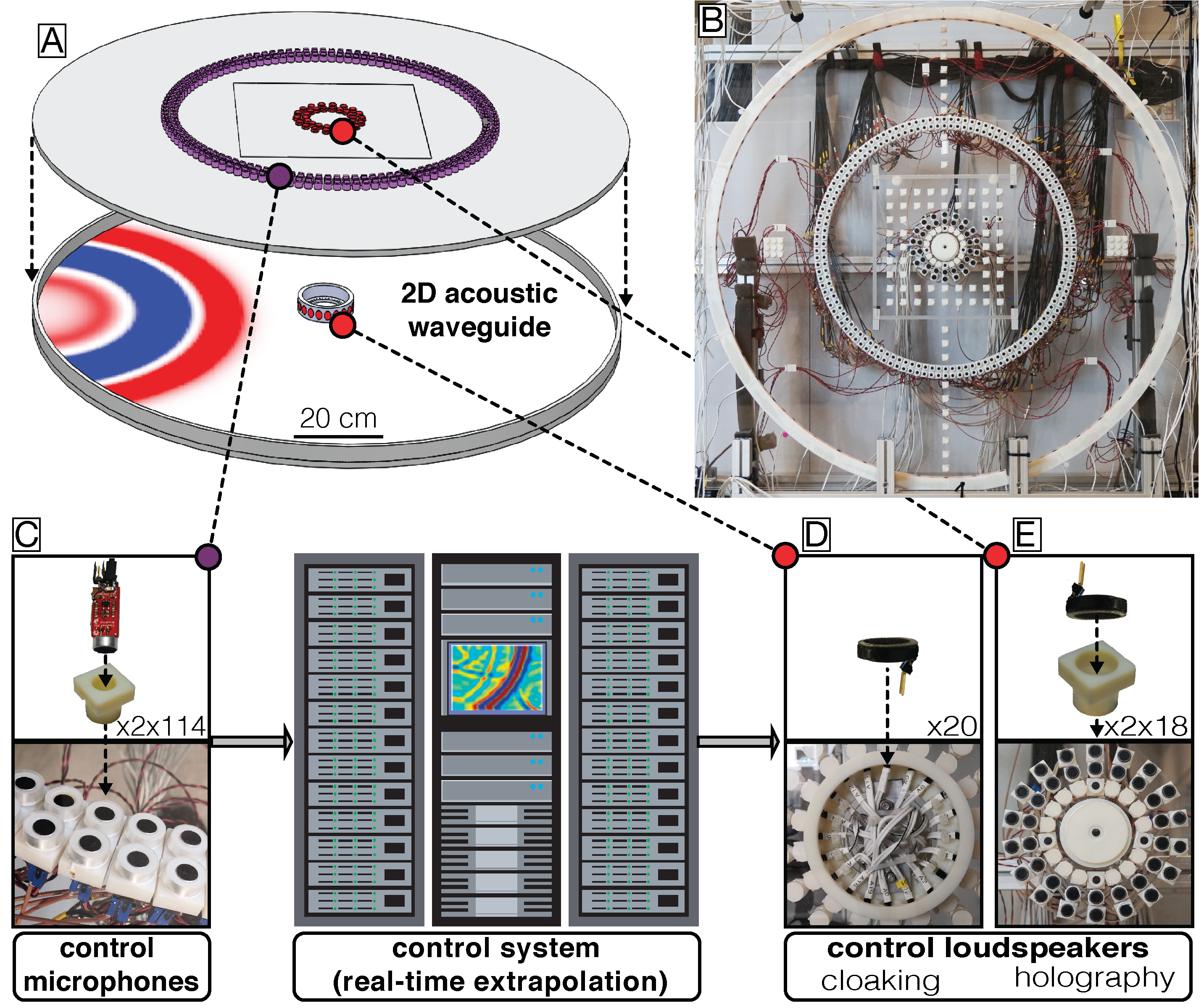}
	\caption{\label{fig:Setup}\baselineskip24pt Schematic (A), photograph (B), and acquisition chain (C-E) for 2D acoustic cloaking and holography experiments. Pressure fields are recorded on two circular arrays with 114 microphones each (C). The recordings are fed to a FPGA-based low-latency control system, which performs the real-time extrapolation to predict the input signals for the control loudspeakers (D~and~E).}
\end{figure}

% FIG 3
\begin{figure}[h!]
	\centering
	\includegraphics{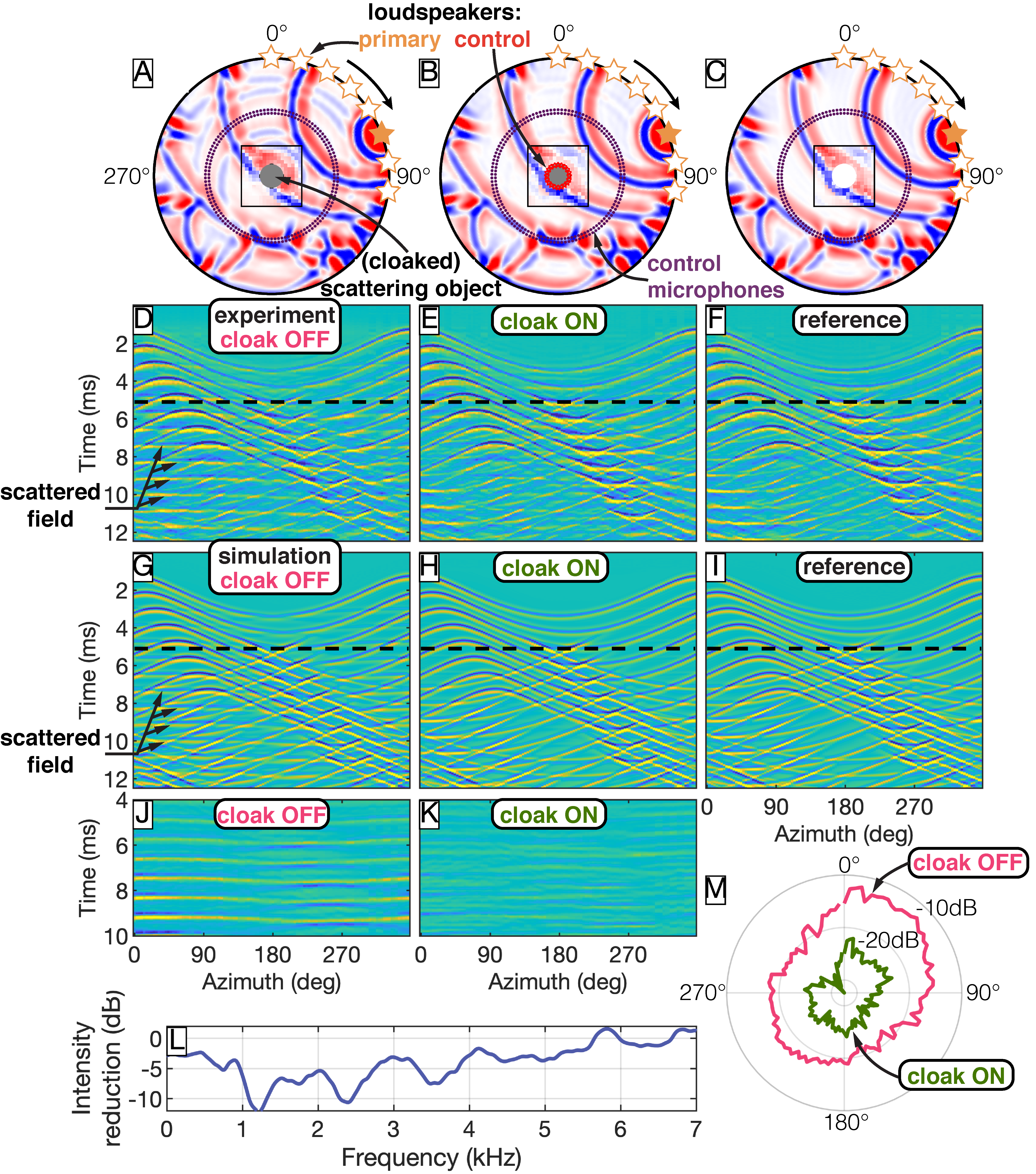}
	\caption{\label{fig:ResultsCloakingRigid}\baselineskip24pt Cloaking of a rigid object. (A-C): Experimental setup and snapshots at 5.1~ms of the experimental data (black square) and corresponding simulations. Experimental and simulated pressure fields at the outer circular microphone array are shown in panels D-F and G-I, respectively. (A,D,G): The object is present inside the waveguide and the cloak is off. (B,E,H): The cloak is switched on; consequently the object is hidden. (C,F,I): References without object inside the waveguide. (J,K): Scattered fields without and with cloak, respectively. (L): Reduction in scattered-field intensity due to active cloaking as a function of frequency. (M): Acoustic intensity in (J) and (K) as a function of angle.}
\end{figure}

% FIG 4
\begin{figure}[h!]
	\centering
	\includegraphics{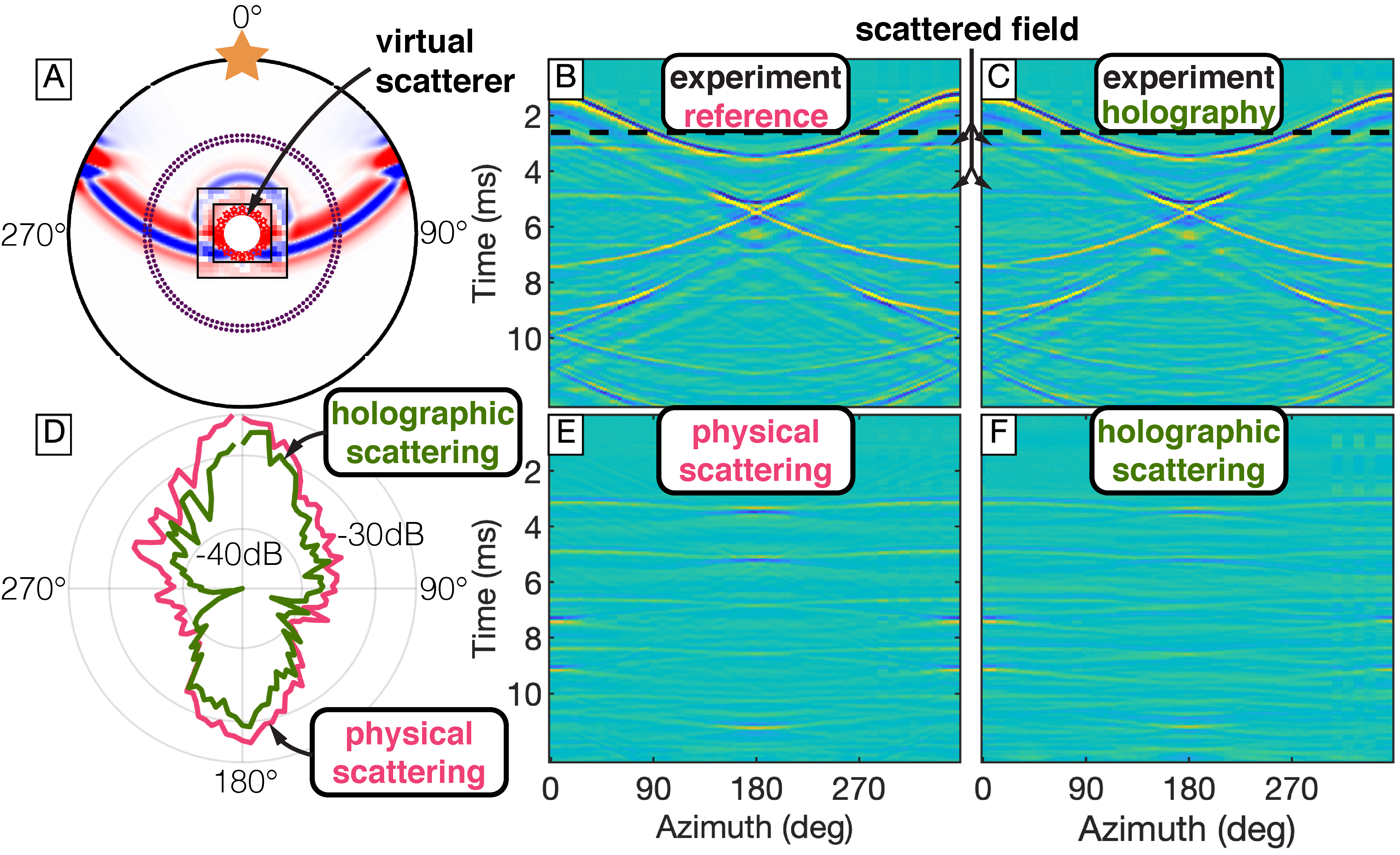}
	\caption{\label{fig:ResultsHolograhpyTransparent}\baselineskip24pt Creation of acoustic holograms at a sound-transparent surface. (A): Experimental setup and snapshots at 2.5~ms of the experimental data (black square) and corresponding simulations. Symbols follow those of Fig.~3%\ref{fig:ResultsCloakingRigid}
. (B): Reference experiment with physical object present inside waveguide. (C): Active creation of a hologram by enabling control loudspeakers (no physical object inside the waveguide). (D): Angular distribution of acoustic intensity in the scattered fields due to a physical (E) and virtual (F) scatterer.}
\end{figure}
\FloatBarrier
\pagebreak
%
%
%
% BEGINNING SUPPLEMENTARY MATERIAL
\section*{Supplementary materials}

%
%\section*{Materials and Methods}
%\subsection*{Experimental Design}
\textbf{Materials and Methods}\\[0.2cm]
\underline{Experimental Design}\\[0.2cm]
An air-filled acoustic waveguide composed of two parallel Polymethyl methacrylate (PMMA) plates with a spacing of $a = 2.5$~cm is used for the presented cloaking and holography experiments. See Fig.~2 %\ref{fig:Setup} 
for photographs of the experimental setup and hardware components. The boundary of the experimental domain consists of a 3D-printed ring with a radius of 66.2~cm made of the synthetic resin VeroWhite. The propagation of waves between the two PMMA plates can be considered 2D as long as signals below the cutoff frequency of the fundamental mode are employed, which is given by $f_\textnormal{c} = 0.5ca^{-1}$ \cite{Redwood1960}, where $c$ is the speed of sound. For our experiments this results in $f_\textnormal{c} \approx 6.9~$kHz. The assumption of 2D wave propagation in the waveguide was validated in previous experiments \cite{Becker2020}. Instead of directly measuring the normal particle velocity and pressure on $S^\textnormal{rec}$, as dictated by the theory of IBCs (see supplementary text below), we employ two circular arrays of pressure microphones with radii of 35.3~cm and 37.3~cm, respectively, each consisting of 114 electret condenser microphones (SparkFun electronics, BOB-12758 and COM-08635, Fig.~2C). The normal particle velocity is then approximated using the pressure gradient in the normal direction\cite{Becker2018, Becker2020a}. All microphones are mounted flush with the inside of the front PMMA plate to minimize their impact on the pressure field. In the case of cloaking experiments, the rigid emitting boundary at $S^\textnormal{emt}$ consists of a 3D-printed ring with a radius of 6.4~cm, housing 20 loudspeakers (PUI Audio, AS01508MR-R, Fig.~2D). For the holography experiments, the transparent emitting boundary at $S^\textnormal{emt}$ consists of two separate circular source arrays with radii of 7.4~cm and 9.4~cm, respectively, each consisting of 18 loudspeakers mounted flush with the front PMMA plate to minimize scattering (Fig.~2E). The distance between $S^\textnormal{rec}$ and $S^\textnormal{emt}$ of approximately 30~cm allows a maximum total latency of about 870~$\mu$s (for experiments in air), of which 200~$\mu$s are needed for the wavefield extrapolation, while the remaining 670~$\mu$s are available for real-time hardware corrections and estimation of normal particle velocity from pressure recordings. The non-local control algorithm is implemented on a massively parallelized, low-latency control system enabled by more than 500 National Instruments FlexRIO field-programmable gate arrays. The system supports simultaneous recording of 800 analogue input channels and simultaneous emission of 800 analogue output channels while operating at a 20~kHz sample rate [more details of the system architecture can be found in \cite{Becker2018}]. 
\\[0.5cm]
\underline{Statistical Analysis and repeatability}\\[0.2cm]
The effectiveness of the broadband active cloak can be assessed and quantified in terms of the reduction in scattered field intensity, $I_\textnormal{R}$, which is calculated according to
\begin{equation}	
	I_\textnormal{R} = \frac{1}{N_r}\sum_{j=1}^{N_r} \left[I_\textnormal{scat,aug}(\bm{x}_j)- I_\textnormal{scat,ini}(\bm{x}_j)\right],
\label{eq:error}
\end{equation}
where $I_\textnormal{scat,aug}(\bm{x}_j)$ and $I_\textnormal{scat,ini}(\bm{x}_j)$ are the acoustic intensities in the cloaked and scattering case, respectively, which are given by 
\begin{equation}	
	I_\textnormal{scat,aug}(\bm{x}_j) = 10\log_{10}\Sigma_{i=1}^{N_t} \hat{p}_\textnormal{scat,aug}^2(t_i,\bm{x}_j)
\label{eq:AIcloaked}
\end{equation}
and
\begin{equation}	
	I_\textnormal{scat,ini}(\bm{x}_j) = 10\log_{10}\Sigma_{i=1}^{N_t} \hat{p}_\textnormal{scat,ini}^2(t_i,\bm{x}_j),
\label{eq:AIscat}
\end{equation}
where $\hat{p}_\textnormal{scat,aug}$ and $\hat{p}_\textnormal{scat,ini}$ are the discrete (residual) scattered pressure fields at measurement locations $\bm{x}_j$ due to the presence of the physical scatterer with and without active cloak, respectively. The total number of recordings and time samples per recording are given by $N_r$ and $N_t$, respectively. The scattered pressure fields $\hat{p}_\textnormal{scat,aug}$ and $p_\textnormal{scat,ini}$ are obtained according to $\hat{p}_\textnormal{scat,aug} = \hat{p}_\textnormal{aug} - \hat{p}_\textnormal{hom}$ and $\hat{p}_\textnormal{scat,ini}  = \hat{p}_\textnormal{ini} - \hat{p}_\textnormal{hom}$, i.e., by subtracting from the total pressure fields the pressure field, $\hat{p}_\textnormal{hom}$, measured in a reference experiment without physical scatterer present. The quantities $\hat{p}_\textnormal{scat,aug} $ and $\hat{p}_\textnormal{scat,ini}$ are displayed in Fig.~3%\ref{fig:ResultsCloakingRigid}
,~K~and~J, respectively, and $I_\textnormal{scat,aug}(\bm{x}_j)$ and $I_\textnormal{scat,ini}(\bm{x}_j)$ are shown in Fig.~3%\ref{fig:ResultsCloakingRigid}
M. A similar expression for reduction in scattered field intensity as a function of frequency can be derived by first transforming $\hat{p}_\textnormal{scat,aug}$ and $\hat{p}_\textnormal{scat,ini}$ to the frequency domain and omitting the summation over time samples (Fig.~3%\ref{fig:ResultsCloakingRigid}
L). 
In the following, we used the recordings of $N_r~=~114$ pressure field microphones on the outer circular microphone array to calculate the reduction in scattered field intensity for the experimental configuration shown in Fig.~3%\ref{fig:ResultsCloakingRigid}
A. The experiments with and without active cloak were repeated a total of 46 times over a duration of approximately 1~hour. This led to a mean $I_\textnormal{R}(\bm{x})$ of -8.4~dB with a standard deviation of 0.05~dB. These findings highlight the strong reduction in scattered field when the cloak is active and also demonstrate the high repeatability of the experiments (see Fig.~S1%\ref{fig:StatsAna}
,~A~and~B). Individual time series from microphones located at azimuths of approximately 0$^{\circ}$, 90$^{\circ}$ and 180$^{\circ}$ are shown in Fig.~S1%\ref{fig:StatsAna}
,~C~to~E. The very good agreement between time series recorded with enabled cloak and time series without scattering object further illustrates the effectiveness of the cloak.\\[0.7cm]
\textbf{Supplementary Text}\\[0.2cm]
\underline{The control algorithm: Immersive boundary conditions}\\[0.2cm]
The control algorithm underlying the experiments of this study is based on immersive boundary conditions (IBCs). IBCs were initially thought-out to obtain non-reflecting domain boundaries in numerical simulations and to embed truncated modeling domains in larger surrounding background media \cite{VanManen2007}. This allows broadband waves from unknown sources to propagate seamlessly between the truncated domain and the background medium, including arbitrary-order scattering interactions due to the recursive nature of the boundary conditions. Ref.~\cite{Vasmel2013} realized that such boundary conditions could also serve as the control algorithm to embed physical wave propagation experiments in surrounding numerical simulations, thereby removing adverse boundary reflections from the experimental domain, overcoming certain wavelength limitations of traditional laboratory experiments, and fully capturing all wave interactions with the numerical background medium. Subsequently, this was demonstrated in 1D and 2D acoustic \cite{Becker2018,Becker2020a} and 1D elastic \cite{Thomsen2019a} experiments, where physical waveguides were embedded in surrounding virtual media. Moreover, Ref.~\cite{VanManen2015} demonstrated in numerical simulations that IBCs can also be employed for cloaking and holography of broadband wavefields, which was experimentally verified in 1D acoustic experiments by \cite{Borsing2019}, for which the underlying wavefield extrapolation reduces to a trivial convolution. In the following, we first summarize the use of IBCs for cloaking of rigid objects and then for creating holograms at a sound-transparent boundary. More detailed derivations can be found in previous studies \cite{VanManen2007,VanManen2015}.
\\[0.2cm] 
\underline{Cloaking of a rigid object}\\[0.2cm]
Consider an object embedded in a homogeneous background medium that scatters an incident wavefield (Fig.~S2%\ref{fig:IBCStates}
A). The pressure field of this \textit{initial} scattering state is denoted with $p_{\textnormal{ini}}$. By emitting appropriate boundary conditions on the surface $S^{\textnormal{emt}}$ enclosing the object, the imprint of the scattering object can be removed from $p_{\textnormal{ini}}$, thereby hiding the object from an observer outside $S^{\textnormal{emt}}$ (Fig.~S2%\ref{fig:IBCStates}
,~B~and~C). The pressure of this \textit{augmented} state in which the object is cloaked is denoted with $p_{\textnormal{aug}}$ and can be understood as the superposition of the pressure in the scattering state and the pressure associated with the wavefield radiated by the boundary conditions, $p_{\textnormal{ibc}}$: 
\begin{equation}	
p_{\textnormal{aug}} = p_{\textnormal{ini}} + p_{\textnormal{ibc}}.
\label{eq:p_aug}
\end{equation}
Hence, the goal of IBCs as the control algorithm is to inject the correct boundary wavefield $p_{\textnormal{ibc}}$ on $S^{\textnormal{emt}}$ to obtain the desired wavefield $p_{\textnormal{aug}}$. The boundary wavefield can be found by differencing the pressure field representations for the initial and the augmented state \cite{VanManen2015}. If the surface $S^\textnormal{emt}$ is aligned with the rigid boundary of the scattering object (i.e., the normal particle velocity on the boundary vanishes), the boundary pressure field is given by \cite{VanManen2015}:
\begin{flalign}
 	 p_\textnormal{ibc,rigid}(\bm{x'},t)  = \oint_{S^\textnormal{emt}}  [G_\textnormal{ini}^{q}(\bm{x}',\bm{x},t) \ast v_{\textnormal{aug},i}(\bm{x},t)]n_i \,dS,
 	 \label{eq:p_ibc_rigid}
\end{flalign}
where $v_{\textnormal{aug},i}(\mathbf{x},t)$ is the $i$-th component of the particle velocity in the augmented (i.e., homogeneous) medium, $G_\textnormal{ini}^{q}(\mathbf{x}',\mathbf{x},t)$ is the acoustic pressure impulse response (Green's functions) of the medium at $\mathbf{x}'$ due to a point source of volume injection rate density, $n_i$ denotes the $i$-th component of the outward-directed normal on $S^\textnormal{emt}$, and an asterisk represents temporal convolution. Equation~(5) %\ref{eq:p_ibc_rigid}
can be interpreted as dense distributions of secondary monopole  sources ($G_\textnormal{ini}^{q}$) on $S^{\textnormal{emt}}$ with source strength $v_{\textnormal{aug},i}(\mathbf{x},t)$. Despite the apparent simplicity of this equation, its implementation in physical real-time experiments poses a significant challenge as the normal particle velocity on $S^\textnormal{emt}$ needs to be known prior to the arrival of the respective waves. To that end, a sound-transparent auxiliary recording surface $S^\textnormal{rec}$ enclosing $S^\textnormal{emt}$ is introduced (Fig.~S2%\ref{fig:IBCStates}
), from which normal particle velocity and pressure wavefields are extrapolated to $S^\textnormal{emt}$ using a Kirchhoff-Helmholtz extrapolation integral \cite{VanManen2015}:
\begin{flalign}
 	v_{\textnormal{aug},i}	 & (\bm{x}^\textnormal{emt},t) \nonumber \\ 
									& \begin{aligned} = \oint_{S^\textnormal{rec}}	& [\Gamma_{\textnormal{aug},i}^{q}(\bm{x}^\textnormal{emt},\bm{x},t) \ast v_{\textnormal{aug},m}(\bm{x},t) \\
																									& + \Gamma_{\textnormal{aug},i,m}^{f}(\bm{x}^\textnormal{emt},\bm{x},t) \ast p_\textnormal{aug}(\bm{x},t)]n_m \,dS. 
 	 \end{aligned}
\label{eq:extrap_v1}
\end{flalign}
Here, $\Gamma_{\textnormal{aug},i}^{q}$ and $\Gamma_{\textnormal{aug},i,m}^{f}$ represent the $i$-th component of the particle velocity impulse response (Green's functions) due to a monopole source and an $m$-directed point-force (dipole) source, respectively, $v_{\textnormal{aug},m}(\bm{x},t)$ and $p_\textnormal{aug}(\bm{x},t)$ are the normal particle velocity and pressure of the augmented state at $S^\textnormal{rec}$, and $n_m$ is the $m$-th component of the outward-pointing normal on $S^\textnormal{rec}$. The Green's functions in Eq.~(6) %\eqref{eq:extrap_v1} 
represent the desired medium inside $S^\textnormal{emt}$, which is a homogeneous volume in the case of cloaking, but in principle, the Green's functions can represent arbitrarily complex, even nonphysical media within $S^\textnormal{emt}$. In that case, an intentional misrepresentations of the actual medium inside $S^\textnormal{emt}$ is obtained. %More details on the extrapolation Green's functions are provided in the Supplementary Text below. 
It is worth noting that the simultaneous extrapolation of particle velocity and pressure implicitly separates the wavefield into ingoing and outgoing components at $S^\textnormal{rec}$ and only the ingoing component is extrapolated to the secondary sources at $S^\textnormal{emt}$, while the outgoing components sum to \textit{zero} \cite{Becker2018}. 
%as $v_{\textnormal{sct},i} \bigg\rvert_{S^\textnormal{emt}}  n_i = 0$ and $G_\textnormal{sct}^{f,i} \bigg\rvert_{S^\textnormal{emt}} n_i= 0$. 
%Similarly, if $S^\textnormal{emt}$ corresponds to a free-surface, Eq.~\ref{eq:p_ibc} reduces to
%\begin{flalign}
% 	 p_\textnormal{ibc,free}(\bm{x'},t)  = \oint_{S^\textnormal{emt}}  [G_\textnormal{sct}^{f,i}(\bm{x}',\bm{x},t)  \ast p_{\textnormal{aug}}(\bm{x},t)]n_i \,dS,
% 	 \label{eq:p_ibc_free}
%\end{flalign}
%as $p_{\textnormal{sct}} \bigg\rvert_{S^\textnormal{emt}} = 0$ and $G_\textnormal{sct}^{q} \bigg\rvert_{S^\textnormal{emt}} = 0$.
\\[0.2cm] 
\underline{Holography at a sound-transparent surface}\\[0.2cm]
In some sense, the creation of holograms can be regarded as the opposite to cloaking: while the initial state is homogeneous, the augmented state contains a desired scattering object (Fig.~S2%\ref{fig:IBCStates}
,~D to~F). The boundary wavefield required to create acoustic holograms can also be found by differencing the pressure field representations in the initial and the augmented states. This time, we do not impose any boundary conditions on $S^\textnormal{emt}$, because in order to create holograms of virtual objects that are not physically present, the emitting surface is, by definition, a transparent surface (Fig.~S2%\ref{fig:IBCStates}
E). In that case the required boundary wavefield is given by \cite{VanManen2015}:
\begin{flalign}
 	 p_\textnormal{ibc,transparent}(\bm{x'},t)  = \oint_{S^\textnormal{emt}}  [G_\textnormal{ini}^{q}(\bm{x}',\bm{x},t) \ast v_i(\bm{x},t) + G_\textnormal{ini}^{f,i}(\bm{x}',\bm{x},t)  \ast p(\bm{x},t) ]n_i \,dS,
 	 \label{eq:p_ibc}
\end{flalign}
with $v_i(\bm{x},t) = v_{\textnormal{aug},i}(\bm{x},t)-v_{\textnormal{ini},i}(\bm{x},t)$ and $p(\bm{x},t) = p_{\textnormal{aug}}(\bm{x},t) - p_{\textnormal{ini}}(\bm{x},t)$, where $v_{\textnormal{aug},i}(\mathbf{x},t)$, $v_{\textnormal{ini},i}(\mathbf{x},t)$, $p_{\textnormal{aug}}(\bm{x},t)$ and $p_{\textnormal{ini}}(\bm{x},t)$ are the $i$-th component of the particle velocity and the pressure in the augmented and initial state, respectively, and $G_\textnormal{ini}^{f,i}(\bm{x}',\bm{x},t)$ is the acoustic pressure impulse response at $\mathbf{x}'$ due to an i-directed force at $\mathbf{x}$. 
Note that, while Eq.~(5) %\ref{eq:p_ibc_rigid} 
only requires monopole sources at the rigid boundary $S^\textnormal{emt}$, the control of waves at a sound-transparent boundary requires monopole and dipole sources with source signatures $v_i(\bm{x},t)$ and $p(\bm{x},t)$, respectively (Fig.~S2%\ref{fig:IBCStates}
E). For that reason, two circular source arrays are used for the holography experiments presented in this study in order to create effective dipole and monopole sources, while for the cloaking experiments a single array suffices. Moreover, for a transparent boundary, the difference of the wavefields in the augmented and initial state are required as the source strengths of the secondary sources on $S^\textnormal{emt}$. As a consequence, four extrapolation integrals (compared to one for a rigid boundary) need to be evaluated to predict the required wavefields on $S^\textnormal{emt}$. However, the number of extrapolation integrals can be reduced to two by differencing the extrapolation Green's functions in the augmented and initial state instead of the extrapolated wavefields:

\begin{flalign}
 	v_{i}	 & (\bm{x}^\textnormal{emt},t) \nonumber \\ 
									& \begin{aligned} = \oint_{S^\textnormal{rec}}	& [\bar{\Gamma}_{i}^{q}(\bm{x}^\textnormal{emt},\bm{x},t) \ast v_{\textnormal{aug},m}(\bm{x},t) \\
																									& + \bar{\Gamma}_{i,m}^{f}(\bm{x}^\textnormal{emt},\bm{x},t) \ast p_\textnormal{aug}(\bm{x},t)]n_m \,dS. 
 	 \end{aligned}
\label{eq:extrap_v2}
\end{flalign}
and
\begin{flalign}
 	p	& (\bm{x}^\textnormal{emt},t) \nonumber \\ 
									& \begin{aligned} = \oint_{S^\textnormal{rec}}	& [\bar{G}^{q}(\bm{x}^\textnormal{emt},\bm{x},t) \ast v_{\textnormal{aug},m}(\bm{x},t) \\
																									& + \bar{G}_{m}^{f}(\bm{x}^\textnormal{emt},\bm{x},t) \ast p_\textnormal{aug}(\bm{x},t)]n_m \,dS, 
 	 \end{aligned}
\label{eq:extrap_p}
\end{flalign}
where, $\bar{\Gamma}_{i}^{q}$, $\bar{\Gamma}_{i,m}^{f}$, $\bar{G}^{q}$ and $\bar{G}_{m}^{f}$ represent the $i$-th component of the \textit{differenced} particle velocity and pressure impulse responses (Green's functions) due to a monopole source and an $m$-directed dipole source, respectively, with $\bar{\Gamma} = \Gamma_{\textnormal{aug}} - \Gamma_{\textnormal{ini}}$ and $\bar{G} = G_{\textnormal{aug}} - G_{\textnormal{ini}}$. This essentially isolates (and extrapolates) the contributions of the virtual scatterer, which are then superimposed on the primary wavefield to create the illusion of a scattering object, but does not extrapolate the direct waves between $S^\textnormal{rec}$ and $S^\textnormal{emt}$. %Another interesting case is obtained by setting $\Gamma_{\textnormal{aug}} = G_{\textnormal{aug}} = 0$, which creates a Nullfield inside $S^\textnormal{emt}$, while not altering the wavefield outside of $S^\textnormal{emt}$. This allows cloaking of arbitrary unknown objects inside $S^\textnormal{emt}$, as no waves propagate within $S^\textnormal{emt}$ \cite{VanManen2015}. 

According to Eqs.~(6),~(8),~and~(9)%\eqref{eq:extrap_v1},~\eqref{eq:extrap_v2},~and~\eqref{eq:extrap_p}
, suitable sets of Green's functions for the extrapolation of particle velocity and pressure from the recording to the emitting surfaces are required to replace the medium within $S^\textnormal{emt}$ with a desired virtual medium. These Green's functions can be analytical, numerically simulated or physically measured [for instance using the approach outlined in \cite{Li2021}]. Here, the Green's functions are obtained by acoustic finite-element modelling of impulsive monopole and dipole sources on $S^\textnormal{emt}$, recording pressure and particle velocity on $S^\textnormal{rec}$, and applying source-receiver reciprocity. The numerical modelling is performed with COMSOL Multiphysics. If, instead, physically recorded scattering Green's functions of an object are used in Eqs.~(8)~and~(9) %\eqref{eq:extrap_v2}~and~\eqref{eq:extrap_p}
, the outlined holography approach would allow to acoustically reproduce the physical object in different physical or virtual environments, thereby acoustically cloning the object.
\\[0.5cm]

\underline{Physical implementation of immersive boundary conditions}\\[0.2cm]
The implementation of immersive boundary conditions in a physical wave propagation laboratory requires overcoming significant practical challenges. As seen in the theory section, the non-local nature of IBCs requires surface-integrals on a mathematically-closed and continuous recording surface to be evaluated in real-time so that control sources on the emitting surface can react appropriately to incoming waves. In our experiments, these continuous surfaces are replaced by dense microphone and loudspeaker arrays. More concretely, measurements from 228 microphones are used for the real-time prediction of the required signals for 20 (cloaking) or 36 (holography) secondary loudspeakers. This constitutes a significant computational effort while requiring an extremely low latency. Hence, a custom-built, massively parallelized data acquisition and control system is used, on which Eqs.~(6),~(8),~and~(9) %\eqref{eq:extrap_v1},~\eqref{eq:extrap_v2},~and~\eqref{eq:extrap_p} 
are implemented by discretizing them (in time and space) and replacing the surface integrals by matrix-vector multiplications. The extrapolation can then be performed recursively at each time step of an experiment \cite{Becker2018}.\\ 
To comply with the theory of IBCs, the physical experiments also require a range of hardware corrections, because ``perfect" measurements of pressure and particle velocity, monopolar/dipolar sources, and ``perfectly rigid boundaries'' are idealized, theoretical constructs that are difficult to achieve in physical experiments. These corrections include the removal of microphone and loudspeaker transfer functions, particle velocity estimation from two parallel microphone arrays, and the creation of effective dipole sources using two parallel loudspeaker arrays. These corrections are represented by scalar operations or short, frequency-dependent matched filters, a few coefficients in length. Due to the associative property of the convolution in Eqs.~(6),~(8),~and~(9)%\eqref{eq:extrap_v1},~\eqref{eq:extrap_v2},~and~\eqref{eq:extrap_p}
, these corrections can be applied to the extrapolation Green's functions prior to an experiment and are then accounted for during the real-time extrapolation of the wavefield. Consequently, the corrections do not require additional computations or filtering operations at run-time. We also attempted to correct for the non-monopolar directivity pattern of the control loudspeakers by spatially filtering the Green's functions according to Ref.~\cite{Li2019}. However, we did not observe a significant decrease in residual errors and hence relied on frequency-dependent corrections only. The hardware corrections and resulting manipulations of the extrapolation Green's functions are summarized in the flowchart in Fig.~S3 %\ref{fig:GFmanip} 
and further details can be found in Refs. \cite{Becker2018,Li2019,Becker2020}.

%
%Start Supplementary Figures
%
%
\renewcommand{\thefigure}{S\arabic{figure}}
\setcounter{figure}{0}   
\clearpage
%
%All Figures and Tables should be cited in order, including those in the Supplementary Material (which should be cited as, for example, “Fig. S1”, and “Table S1”). 

\begin{figure}[h!]
	\centering
	\includegraphics{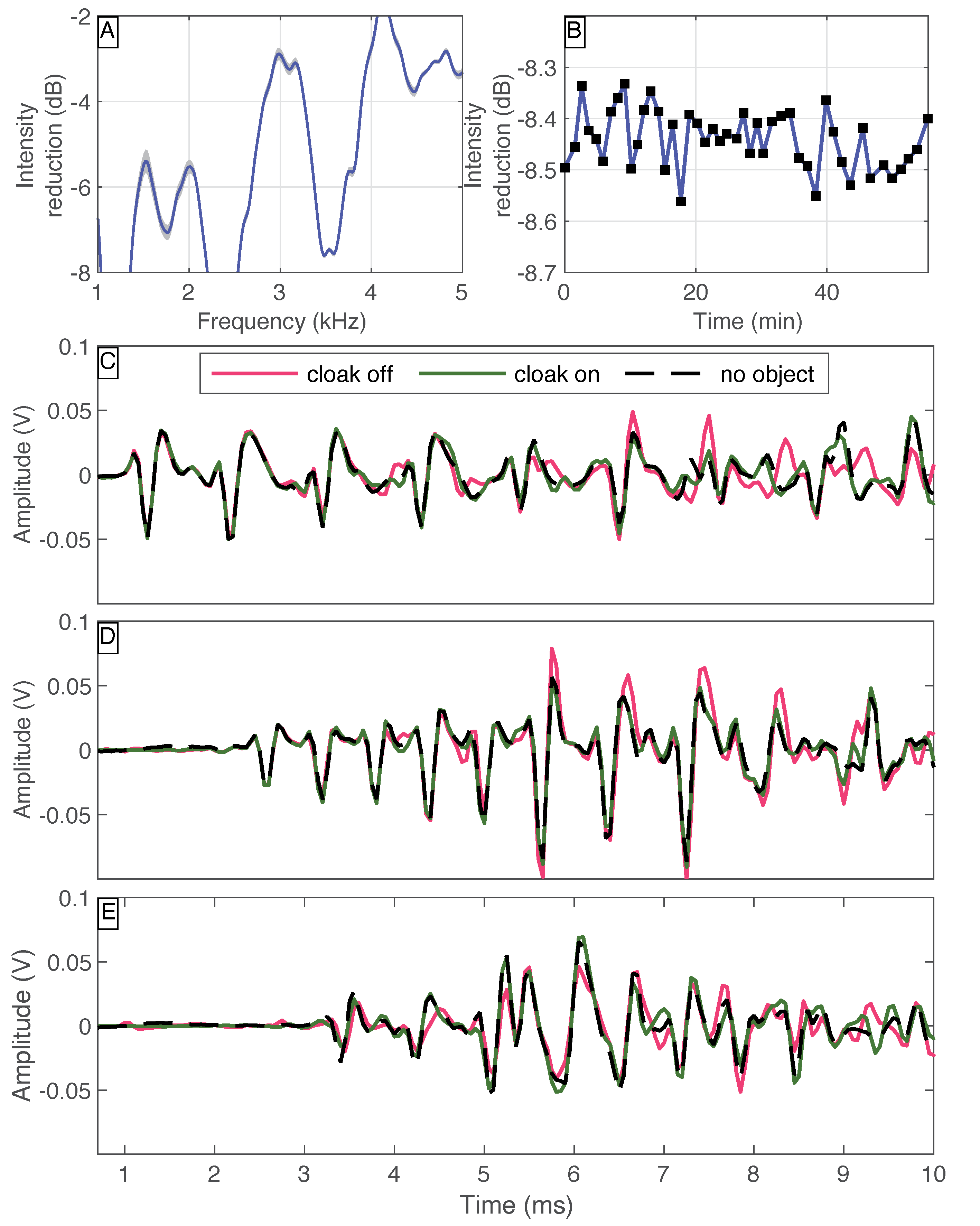}
	\caption{\label{fig:StatsAna}Error analysis for active broadband cloaking experiments. (A): mean intensity reduction of the scattered field with enabled cloak for 46 realizations of the same cloaking experiment (blue line) and the according standard deviation (grey area). (B): Intensity reduction of the same 46 realizations over time. (C-E): Times series measured with microphones on the outer circular array at azimuths of approximately 0$^{\circ}$, 90$^{\circ}$ and 180$^{\circ}$. Note that the green and black dashed lines largely overlay due to the good agreement between augmented and reference results.}
\end{figure}
\clearpage

\begin{figure}[h!]
	\centering
	\includegraphics{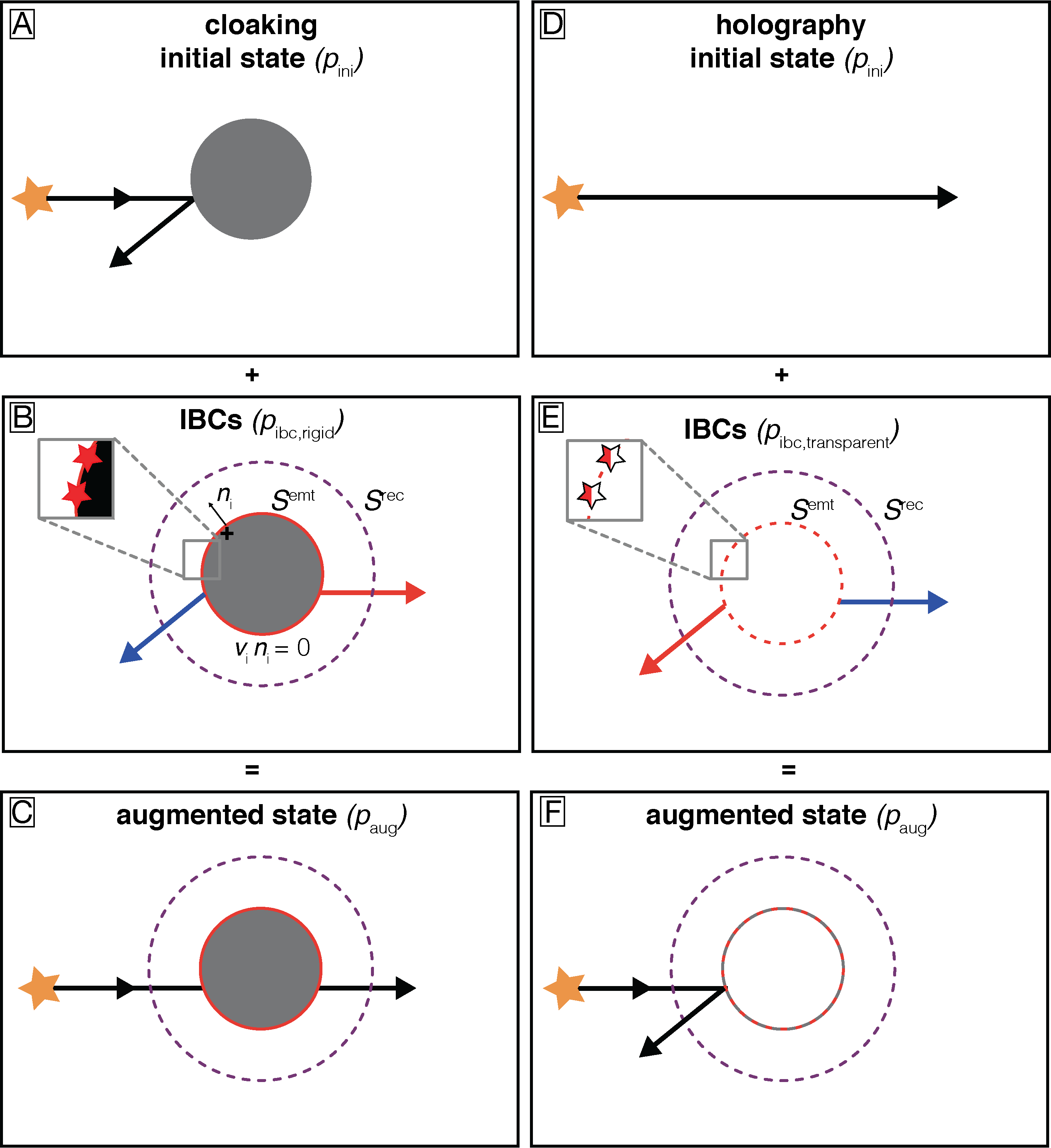}
	\caption{\label{fig:IBCStates}Illustration of wavefields for cloaking (A-C) and holography (D-F). (A): The wavefield of a primary source (orange star) is scattered by a rigid object (black circle) yielding the initial wavefield $p_{\textnormal{ini}}$. (B): The boundary wavefield, $p_{\textnormal{ibc,rigid}}$, is emitted by monopole sources on the rigid boundary $S^\textnormal{emt}$ (red solid line / red stars). This boundary wavefield interacts with the primary field such that the reflected waves are canceled (blue arrow) and the transmitted waves are reproduced (red arrow). The signatures of the monopole sources are obtained by forward extrapolation from the sound-transparent recording boundary $S^\textnormal{rec}$ (dashed purple line). (C): The superposition of wavefields $p_{\textnormal{ini}}$ and $p_{\textnormal{ibc,rigid}}$ yields the augmented state, $p_{\textnormal{aug}}$, for which the rigid scatterer is rendered invisible. For holography, an initially undisturbed wavefield propagating in a homogeneous medium (D) is augmented by the emission of collocated monopole and dipole sources on the (now) sound-transparent surface $S^\textnormal{emt}$ (E, red dashed line / white-red stars) to produce the forward-scattered (blue arrow) and back-scattered (red arrow) fields of a virtual object (E and F).}
\end{figure}
\clearpage

\begin{figure}[h!]
	\centering
	\includegraphics{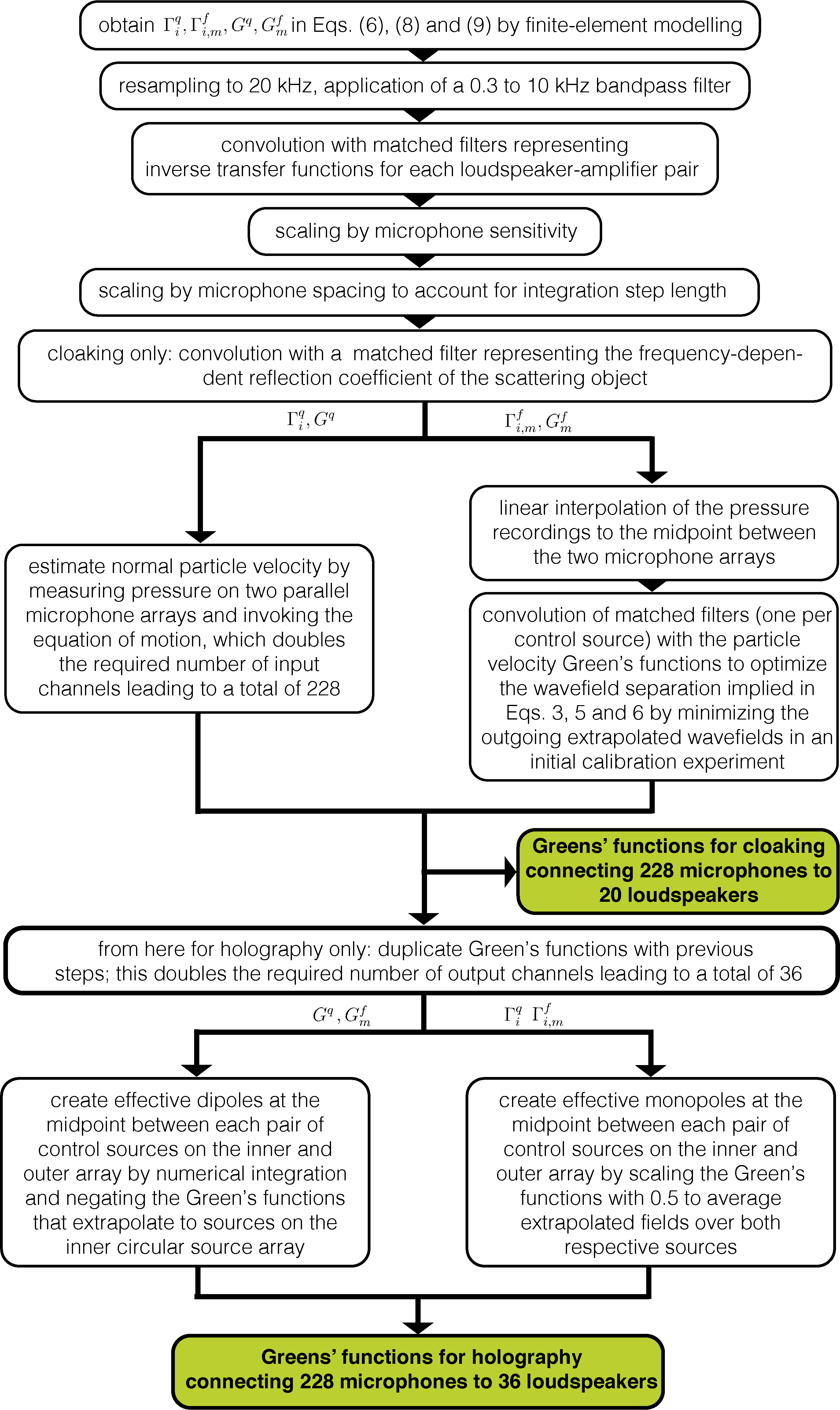}
	\caption{\label{fig:GFmanip}Flowchart describing the generation and processing of the extrapolation Green's functions for cloaking and holography experiments. For holography experiments, the workflow is repeated twice: once for Green's functions of the augmented state and once for Green's functions of the initial state.}
\end{figure}
\clearpage

\end{document}